\documentclass{article}

\usepackage{arxiv}

\usepackage[utf8]{inputenc} 
\usepackage[T1]{fontenc}    
\usepackage{hyperref}       
\usepackage{url}            
\usepackage{booktabs}       
\usepackage{amsfonts}       
\usepackage{nicefrac}       
\usepackage{microtype}      
\usepackage{lipsum}		
\usepackage{graphicx}
\usepackage[numbers]{natbib}
\usepackage{doi}
\usepackage{comment}

\title{AmBox: Device-to-Blockchain\\Ambient Sensing for Food Traceability}

\author{ \href{https://orcid.org/0009-0005-5650-0463}
    {\includegraphics[scale=0.06]{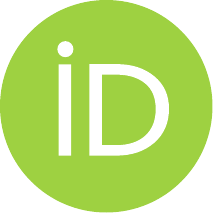}
   \hspace{1mm}João Miguel Guerreiro Fernandes} \\
	  INESC-ID, Instituto Superior Técnico,\\Universidade de Lisboa, Portugal.\\
    \texttt{joao.m.g.fernandes@tecnico.ulisboa.pt} \\
	  \And
	\href{https://orcid.org/0000-0003-0972-4171}
    {\includegraphics[scale=0.06]{orcid.pdf}
    \hspace{1mm}Samih Eisa} \\
    INESC-ID, Instituto Superior Técnico,\\Universidade de Lisboa, Portugal.\\
	\texttt{samih.eisa@inesc-id.pt} \\
    \And
	\href{https://orcid.org/0000-0003-2872-7300}
    {\includegraphics[scale=0.06]{orcid.pdf}
    \hspace{1mm}Miguel L. Pardal} \\
    INESC-ID, Instituto Superior Técnico,\\Universidade de Lisboa, Portugal.\\
	\texttt{miguel.pardal@tecnico.ulisboa.pt} \\
}

\renewcommand{\shorttitle}{AmBox}

\hypersetup{
pdftitle={AmBox: Device-to-Blockchain Ambient Sensing for Food Traceability},
pdfsubject={Food traceability, ambient sensing, blockchain, Hyperledger Fabric},
pdfauthor={João Miguel Guerreiro Fernandes, Samih Eisa, Miguel L. Pardal},
pdfkeywords={food traceability, ambient sensing, blockchain, device-to-blockchain, Hyperledger Fabric, Raspberry Pi, ESP32, supply chain, temperature monitoring, humidity monitoring},
}

\begin{document}
\maketitle

\begin{abstract}
From production to consumption, ensuring food quality and traceability depends on reliable monitoring of environmental conditions across the supply chain. Ambient sensing devices can collect relevant data such as temperature and humidity, but ensuring its integrity among stakeholders remains a challenge. This work presents \emph{AmBox}, a system that enables device-to-blockchain ambient sensing for food traceability. \emph{AmBox} connects sensors to a blockchain, ensuring secure, verifiable, and tamper-resistant data collection with minimal intermediaries. It manages sensor commissioning and operation with the adequate business context. \emph{AmBox} can operate with standalone nodes or within a distributed node-mote architecture, allowing flexible deployment at different points along the supply chain. A prototype using Raspberry Pi and ESP32 hardware can record sensor data directly on Hyperledger Fabric. Experimental results show that \emph{AmBox} provides timely and reliable data that can increase transparency and trust between the supply chain stakeholders.
\end{abstract}


\section{Introduction}

The agri-food supply chain is essential for sustaining human life and supporting global economic development. 
It encompasses multiple stages, ranging from crop cultivation and harvesting to processing, storage, transportation, and final delivery to consumers~\cite{Tsang2019}. 
Managing perishable goods, such as fresh produce, is particularly challenging due to their limited shelf life and high sensitivity to environmental conditions. 
Inadequate handling or monitoring can lead to product spoilage, food safety risks, increased waste, and a loss of consumer trust.

Recent advances in Internet of Things (IoT) technologies and data analytics have significantly improved the feasibility of real-time supply chain monitoring~\cite{Misra2022}. 
Low-cost and energy-efficient sensors can now be deployed across diverse environments, including agricultural fields, transportation routes, and indoor storage or processing facilities. 
The data collected from these sensors enables continuous visibility into operational conditions and supports informed decision-making throughout the supply chain.

Agri-food supply chains typically involve multiple independent organizations, each maintaining its own records and information systems. 
This fragmentation can lead to data inconsistencies, limited transparency, and disputes between stakeholders. 
Establishing a shared and trustworthy data layer is therefore important to ensuring data consistency, accountability, and trust across all participants.

Blockchain technology offers promising capabilities for addressing these challenges by providing a decentralized, tamper-resistant, and auditable data infrastructure~\cite{10.3390/su12083497}. 
In a blockchain-based system, records are stored immutably and shared among participants without reliance on a centralized trusted third party. 
For supply chain applications, permissioned blockchains, such as Hyperledger Fabric~\cite{androulaki2018hyperledger}, are particularly suitable, as participants are known, authenticated, and authorized in advance. 
These systems typically rely on Proof-of-Authority (PoA) consensus mechanisms, which offer higher efficiency and lower energy consumption compared to mining- or staking-based approaches used in public blockchains~\cite{Xiao2019ASO}.

Once such a blockchain infrastructure is in place, an important challenge is enabling sensing devices to upload trustworthy data to the ledger in a timely manner and with minimal intermediaries. 
This is especially relevant in environments with constrained connectivity, where intermittent network access and energy limitations are common.

In this paper, we propose \emph{AmBox} (Ambient Monitoring Box), a system that integrates IoT sensing and blockchain technologies to enhance traceability, transparency, and food-safety monitoring in agri-food supply chains. 
\emph{AmBox} enables direct and auditable recording of environmental data from the source and across multiple stages of storage and transportation, reducing reliance on third-party intermediaries and increasing trust among stakeholders.

We develop a functional prototype of the \emph{AmBox} system and evaluate its applicability in realistic scenarios. 
The prototype monitors key environmental parameters, including temperature, humidity, and atmospheric pressure, that are critical for preserving the quality of perishable goods.

The system is evaluated through a series of experiments designed to emulate real-world operating conditions, with a particular focus on the cherry supply chain. 
Cherries are highly perishable fruits with a short shelf life and strict environmental requirements~\cite{10.24326/asphc.2022.6.3}, making them a representative and demanding use case. 
The evaluation focuses on communication reliability, energy consumption, sensor accuracy, and the system’s robustness to network disruptions, all of which are essential for ensuring product quality throughout the distribution process.

The main contributions of this work are as follows:
\begin{itemize}
    \item The design of a low-cost, modular, and portable system for environmental monitoring in agri-food supply chains, capable of operating under constrained connectivity while ensuring data reliability through blockchain-based auditing;
    \item The development of two complementary hardware prototypes: \emph{AmBox Node}, a Raspberry Pi-based sensing device with integrated environmental sensors and blockchain connectivity; and \emph{AmBox Mote}, an ESP32-based auxiliary device equipped with external sensors and local storage, enabling delayed data forwarding when network access is unavailable;
    \item An experimental evaluation using controlled scenarios to assess sensor accuracy, communication latency, energy consumption, and the system’s resilience to intermittent connectivity.
\end{itemize}

\section{Background}
This section reviews the key concepts and technological foundations underlying the development of the \emph{AmBox} system, focusing on supply chains, blockchain technology, and the Internet of Things.

\subsection{Supply Chain}

A supply chain is a network of organizations that coordinate activities related to the production, processing, storage, transportation, and distribution of products to end consumers \cite{Mentzer2001}. 
In the agri-food sector, supply chains are characterized by multiple stakeholders, geographically distributed operations, and strict requirements related to food safety and quality.

Growing concerns regarding consumer health, environmental sustainability, and regulatory compliance have intensified the demand for improved traceability and transparency in agri-food supply chains. 
Accurate and timely access to data describing product origin, handling conditions, and logistics processes is important for ensuring accountability and maintaining consumer trust. 
To address these challenges, digital technologies, particularly blockchain and Internet of Things (IoT) solutions, are increasingly adopted to collect, share, and verify supply chain data in a reliable and auditable manner \cite{10.1016/j.jclepro.2020.121031}.

\subsection{Blockchain}

Blockchain is a distributed ledger technology that enables secure, transparent, and tamper-resistant storage of data through a consensus-driven, append-only structure \cite{10.1109/jiot.2019.2920987}. 
By replicating the ledger across multiple nodes, blockchain eliminates the need for a centralized trusted authority while ensuring that recorded data cannot be altered retroactively.

Permissioned blockchains, such as Hyperledger Fabric \cite{androulaki2018hyperledger}, are particularly suitable for enterprise and supply chain applications. 
In these systems, participation is restricted to authenticated entities, allowing for controlled access, improved performance, and compliance with organizational and regulatory requirements. 
Hyperledger Fabric further supports modular components, private channels, and fine-grained access control, enabling selective data sharing among stakeholders.

Smart contracts, also referred to as chaincode in Hyperledger Fabric, are executable programs deployed on the blockchain that automate business logic and enforce predefined rules in a transparent and verifiable manner \cite{Christidis2016}. 
They play a key role in coordinating supply chain processes and ensuring consistent data handling across participants.

\subsection{Internet of Things}

The Internet of Things (IoT) refers to a network of interconnected physical objects equipped with sensors, software, and communication capabilities that enable autonomous data collection, exchange, and processing \cite{Talwana2016}. 
IoT systems are widely used to support monitoring and automation across domains such as agriculture, logistics, industrial production, and smart environments.

In agri-food supply chains, IoT devices are commonly deployed to monitor environmental conditions, track product movement, and provide real-time visibility into storage and transportation processes. 
These capabilities are particularly important for perishable goods, where environmental deviations can rapidly affect product quality and safety.

\subsubsection{Devices and Power}

IoT devices typically rely on either microcontrollers or single-board computers, depending on the computational requirements of the application. 
Microcontroller-based platforms are optimized for low-power operation and are commonly used in battery-powered and resource-constrained scenarios, while single-board computers, such as the Raspberry Pi, provide greater processing capabilities and support more complex software stacks \cite{10.1007/978-3-030-72369-9_9}.

Power supply considerations play a critical role in IoT system design. 
Devices may operate on batteries, suitable for mobile or intermittently connected deployments, or rely on continuous power sources in fixed installations. 
Balancing energy consumption with sensing, processing, and communication requirements is essential for ensuring reliable long-term operation.

\subsubsection{Communication}

IoT communication relies on a variety of wireless technologies, including Wi-Fi, Bluetooth, Zigbee, and long-range solutions such as LoRa, each offering different trade-offs in terms of bandwidth, energy consumption, and coverage \cite{Devalal2018}. 
At the application layer, communication protocols such as MQTT, CoAP, HTTP, and gRPC are commonly used to exchange data between devices, gateways, and backend services.

Depending on the deployment context, IoT devices may experience continuous, periodic, or delay-tolerant network connectivity \cite{Kumar2023}. 
In supply chain scenarios, particularly during transportation, connectivity disruptions are common, requiring systems to support local data storage and deferred data transmission to ensure reliable data collection and delivery.

\section{Related Work}

The integration of blockchain and Internet of Things (IoT) technologies has been widely investigated as a means to improve traceability, transparency, and trust in supply chain systems. 
Several studies have specifically focused on agri-food supply chains, where data integrity and environmental monitoring are essentials.

A number of works propose blockchain-based architectures for validating and storing sensor data collected from IoT devices. 
Moudoud \emph{et al.} introduced \emph{LC4IoT}, a lightweight blockchain architecture that relies on oracles and a four-tier validation model to ensure the reliability of sensor data before recording it on the blockchain \cite{Moudoud2019}. 
Balamurugan \emph{et al.} proposed a traceability approach based on QR codes to track food products along the supply chain, aiming to improve data transparency and security \cite{Balamurugan2021}. 
Cocco \emph{et al.} implemented a decentralized bread traceability system combining RFID technology, smart contracts, and cloud-based storage to record production and distribution events \cite{Cocco2021}. 
Khan \emph{et al.} explored the use of private blockchain platforms together with deep learning models, such as Long Short-Term Memory (LSTM) and Gated Recurrent Units (GRU), to enhance supply chain security and enable demand forecasting and sales prediction \cite{Khan2020}.

In parallel, several studies have focused on the design of ambient and environmental monitoring devices for food logistics and cold-chain management. 
Sarkar \emph{et al.} proposed a cold-chain monitoring device equipped with sensors for gas concentration, temperature, humidity, and light exposure, enabling detailed tracking of storage and transportation conditions \cite{Sarkar2022}. 
Misra \emph{et al.} investigated the use of UV-Vis-NIRS sensors for food fingerprinting, demonstrating how spectroscopic sensing combined with chemometric analysis can support quality assessment and authentication \cite{Misra2022}. 
Other works advocate for the combined use of artificial intelligence, IoT, and blockchain technologies to enhance food safety, quality monitoring, and traceability through advanced sensing and data analytics techniques \cite{Bhat2021}.

While these approaches demonstrate the potential of combining IoT, blockchain, and data analytics in agri-food supply chains, many existing solutions rely on centralized cloud infrastructures, assume continuous network connectivity, or require complex and costly sensing equipment. 
Moreover, limited attention is often paid to the practical constraints of real-world deployments, such as energy efficiency, intermittent connectivity, and low-cost hardware integration with permissioned blockchain platforms. 
In contrast, the \emph{AmBox} system focuses on providing a lightweight, modular, and resilient monitoring solution that directly integrates IoT sensing devices with a permissioned blockchain, supporting delayed data submission and reliable operation under constrained connectivity conditions.

\section{AmBox Design}

\emph{AmBox} is a compact and low-cost system designed to monitor environmental conditions in agri-food supply chains. 
Unlike traditional monitoring solutions that rely on centralized servers or trusted intermediaries, \emph{AmBox} records monitoring data directly onto a blockchain, ensuring tamper resistance, decentralization, and transparent data sharing among stakeholders. 
The system is intended for deployment in storage facilities and transport vehicles, where it continuously monitors ambient conditions such as temperature and humidity that are essential to the preservation of perishable goods.

\subsection{Requirements}

The design of \emph{AmBox} is guided by the following requirements:

\begin{itemize}
    \item \textbf{Direct blockchain interaction:} \emph{AmBox} devices must be capable of interacting directly with blockchain peers through smart contracts, without relying on external gateways, centralized servers, or third-party intermediaries;
    \item \textbf{Connectivity resilience:} Devices must operate autonomously during extended periods of limited or absent network connectivity, locally buffering monitoring data and securely uploading it once connectivity is restored;
    \item \textbf{Tamper-evident communication:} All transmitted monitoring data must be protected using cryptographic mechanisms such that any modification during storage or transmission can be detected;
    \item \textbf{Energy autonomy:} Devices must operate independently of external power sources for prolonged periods. Based on the identified deployment scenarios, each device should achieve a minimum battery life of 10 hours under typical operating conditions;
    \item \textbf{Scalability:} The system must support deployments ranging from individual transport units to large-scale warehouse environments, allowing multiple devices to be easily added, configured, and managed within the same network.
\end{itemize}

\subsection{Architecture}

\begin{figure}[!htb]
    \centering
    \includegraphics[width=0.8\columnwidth]{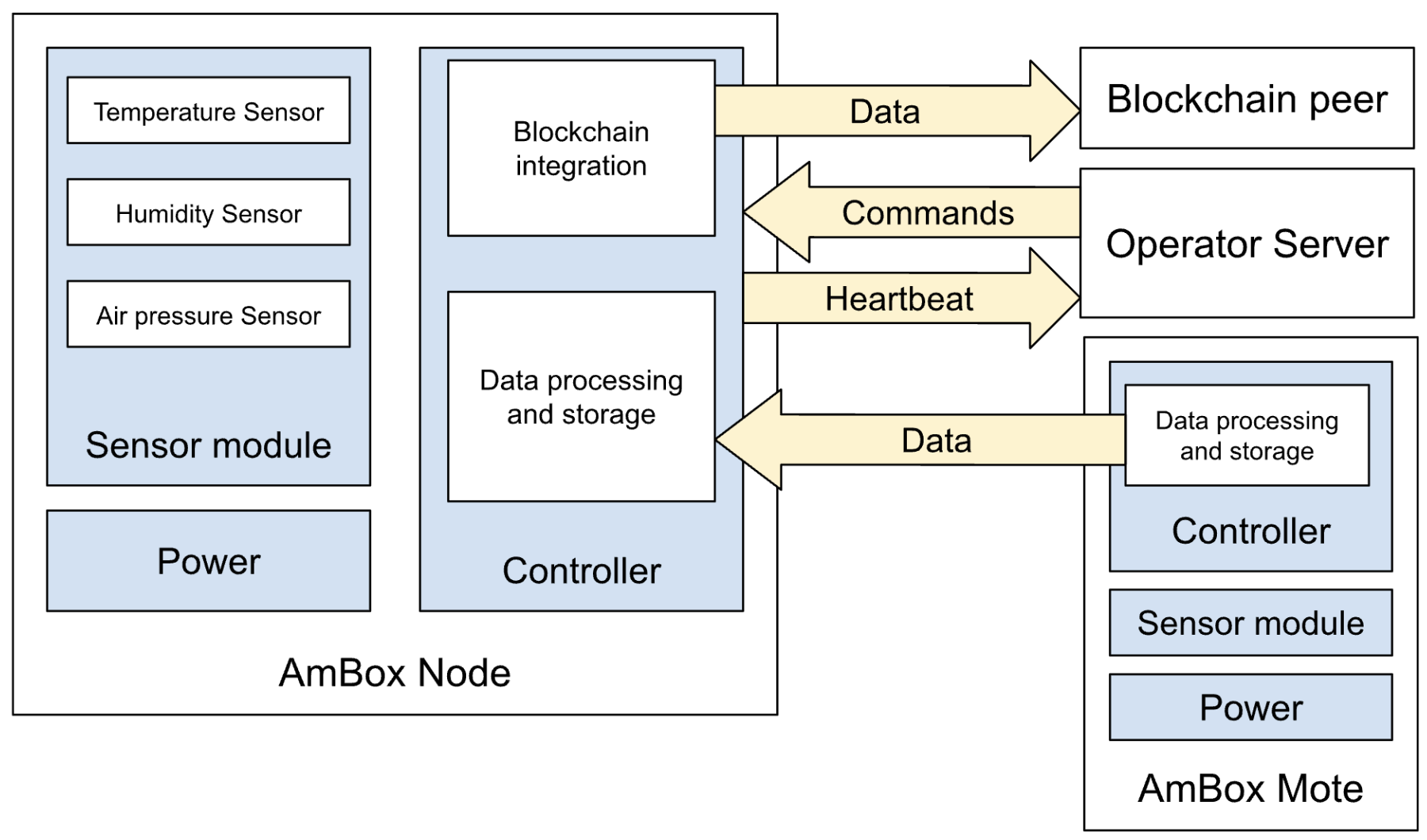}
    \caption{AmBox system architecture}
    \label{fig:ambox-architecture}
\end{figure}

Figure~\ref{fig:ambox-architecture} presents the overall architecture of the \emph{AmBox} system. 
A dedicated sensing module is responsible for acquiring environmental data from the physical environment, while a controller module handles data processing, local storage, and communication with the blockchain. 
The architecture follows a modular and scalable design organized into three logical layers.

\subsubsection{Sensing Layer}

The sensing layer is responsible for capturing environmental parameters such as temperature, humidity, and atmospheric pressure. 
It can also integrate optional modules, including GPS receivers, and aggregate data from nearby devices.

Two types of devices operate at this layer:

\begin{itemize}
    \item \textbf{AmBox Node:} The Node is the primary device in the system. 
    It collects sensor data, performs local processing, stores data during offline periods, and submits signed records to the blockchain. 
    A Node can operate independently or act as a hub for multiple Motes. 
    It supports offline operation, digital signatures, and energy-efficient communication.
    
    \item \textbf{AmBox Mote:} Motes are low-power sensing devices designed to increase spatial monitoring granularity. 
    They collect environmental data and transmit it to a nearby Node using short-range communication technologies. 
    Motes do not interact with the blockchain directly and rely on the Node to forward their data securely.
\end{itemize}

\begin{figure}[!htb]
    \centering
    \includegraphics[width=0.6\columnwidth]{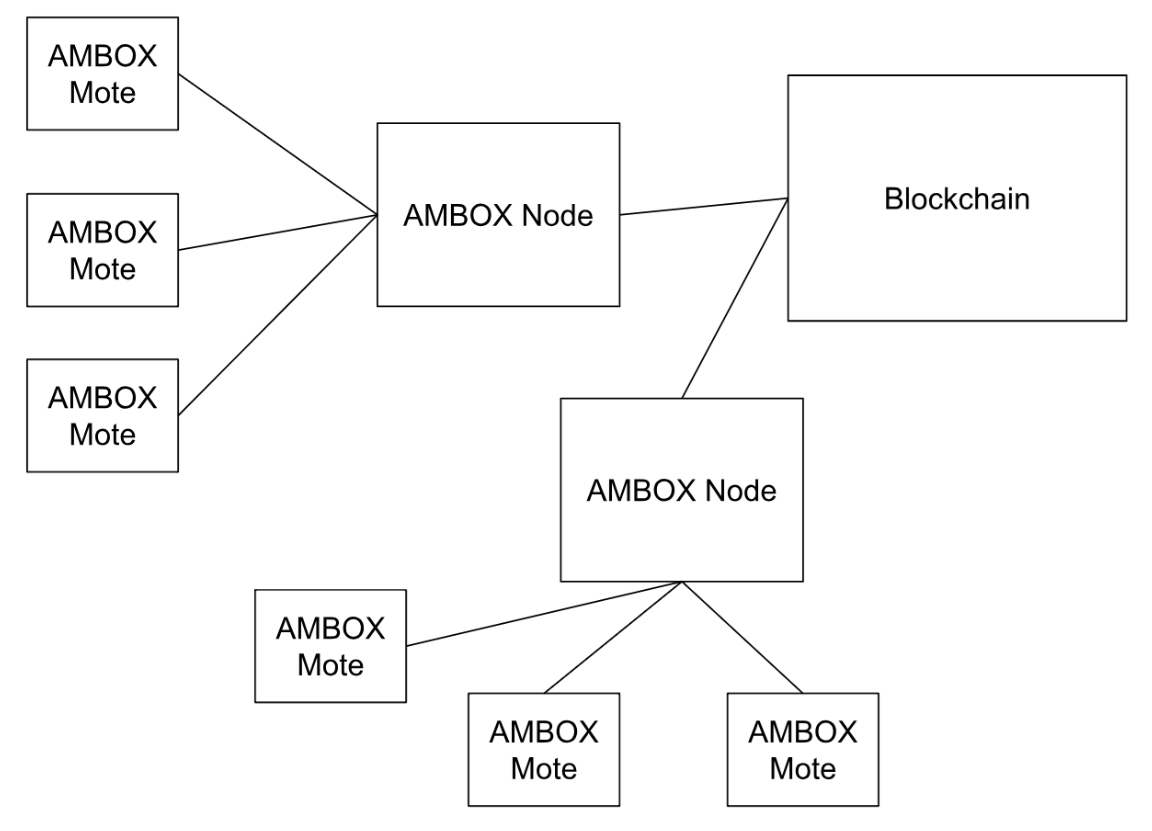}
    \caption{Multi-device deployment with AmBox Nodes and Motes}
    \label{fig:architecture-multi-device}
\end{figure}

Depending on the deployment scenario, the system can operate in two modes: a single-device configuration, in which an \emph{AmBox Node} performs both sensing and blockchain communication, and a multi-device configuration, in which a Node aggregates data from multiple \emph{AmBox Motes}. 
This hierarchical organization enables flexible and scalable deployments while minimizing energy consumption at the sensing edge \cite{Weyns2017}. 
Figure~\ref{fig:architecture-multi-device} illustrates the interaction between Nodes and Motes in a distributed sensing setup.

\subsubsection{Data Processing and Storage Layer}

The data processing and storage layer manages local data handling within the Node. 
It aggregates data received from onboard sensors and Motes, applies preprocessing when required, and securely stores records during periods of network unavailability. 
All stored data is digitally signed to ensure integrity and non-repudiation prior to blockchain submission.

\subsubsection{Blockchain Integration Layer}

The blockchain integration layer is responsible for submitting signed monitoring records to the blockchain via smart contracts. 
This layer ensures immutable logging, traceability, and secure access to monitoring data for authorized stakeholders. 
By directly interfacing with the blockchain network, the system avoids reliance on centralized intermediaries while maintaining auditability and transparency.

\subsection{Lifecycle}

To enhance reliability and enable remote supervision, each device periodically emits a \textit{heartbeat} message that indicates its operational status during idle periods.

\begin{figure}[!htb]
    \centering
    \includegraphics[width=0.6\columnwidth]{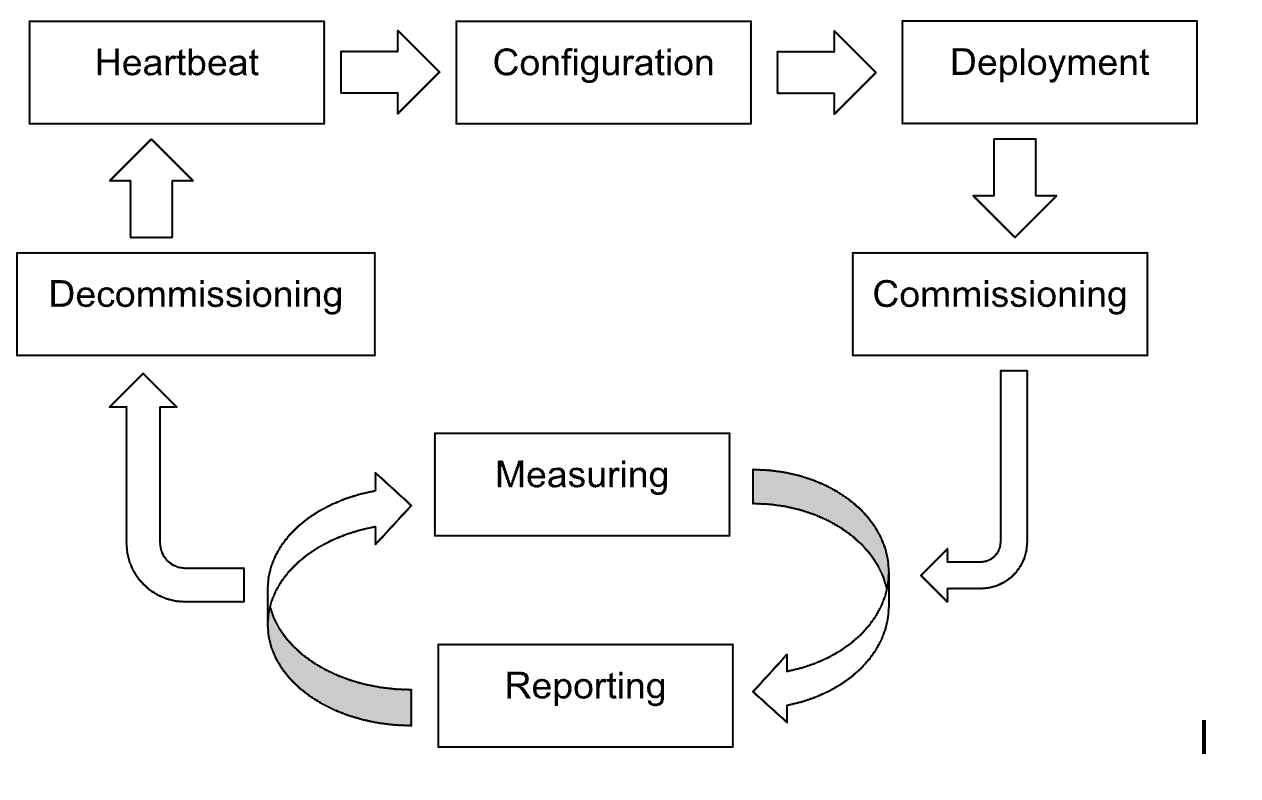}
    \caption{AmBox lifecycle}
    \label{fig:AmBox-lifecycle}
\end{figure}

The \emph{AmBox} lifecycle spans from initial configuration to decommissioning, as illustrated in Figure~\ref{fig:AmBox-lifecycle}. 
The process begins with a configuration phase, during which device parameters, credentials, and monitoring policies are defined. 
Following deployment and commissioning, the device enters the operational phase, where it performs continuous sensing and reports data to the blockchain according to the configured schedule.

When a decommissioning command is issued, the device terminates active monitoring and transitions to a heartbeat-only stage, in which it periodically reports its status to the operator infrastructure, enabling verification of proper shutdown and traceability of device usage.

\section{System Features}

This section describes the key functional and non-functional features of the \emph{AmBox} system.
These features capture the system’s core capabilities and design-level guarantees, independently of the specific hardware and software technologies used for implementation.

\subsection{Concurrency and Fault Tolerance}

The \emph{AmBox} system supports concurrent sensor operations by assigning independent data collection loops to each sensor. 
Sensor data is first stored locally in JSON format, enabling continued operation during temporary network or blockchain outages. 
Failed transmissions are retried automatically, ensuring continuity of operation and minimizing data loss.

\subsection{Data Integrity}

Each monitoring record is digitally signed using the RSA-SHA256 algorithm before submission to the blockchain. 
The corresponding public key is used to verify the signature on-chain. 
Any modification to the data after signing invalidates the signature, providing strong guarantees of integrity and authenticity.

\subsection{Scalability and Flexibility}

\emph{AmBox} is implemented using a modular architecture that supports both standalone and multi-device deployments. 
A single Node can aggregate data from multiple Motes, while multiple Nodes can operate independently and submit data to the blockchain in parallel. 
This design enables horizontal scaling and flexible deployment across different stages of the agri-food supply chain.

\section{AmBox Implementation}

This section describes the implementation of the \emph{AmBox} system, detailing the hardware and software components used to realize the proposed design, as well as the communication mechanisms between system entities.

\subsection{Hardware Setup}

The \emph{AmBox} system consists of four main hardware and infrastructure components: the \emph{AmBox Node}, the \emph{AmBox Mote}, the Operator Server, and the Blockchain Network.

\subsubsection{AmBox Node}

\begin{figure}[!htb]
    \centering
    \includegraphics[width=0.7\linewidth]{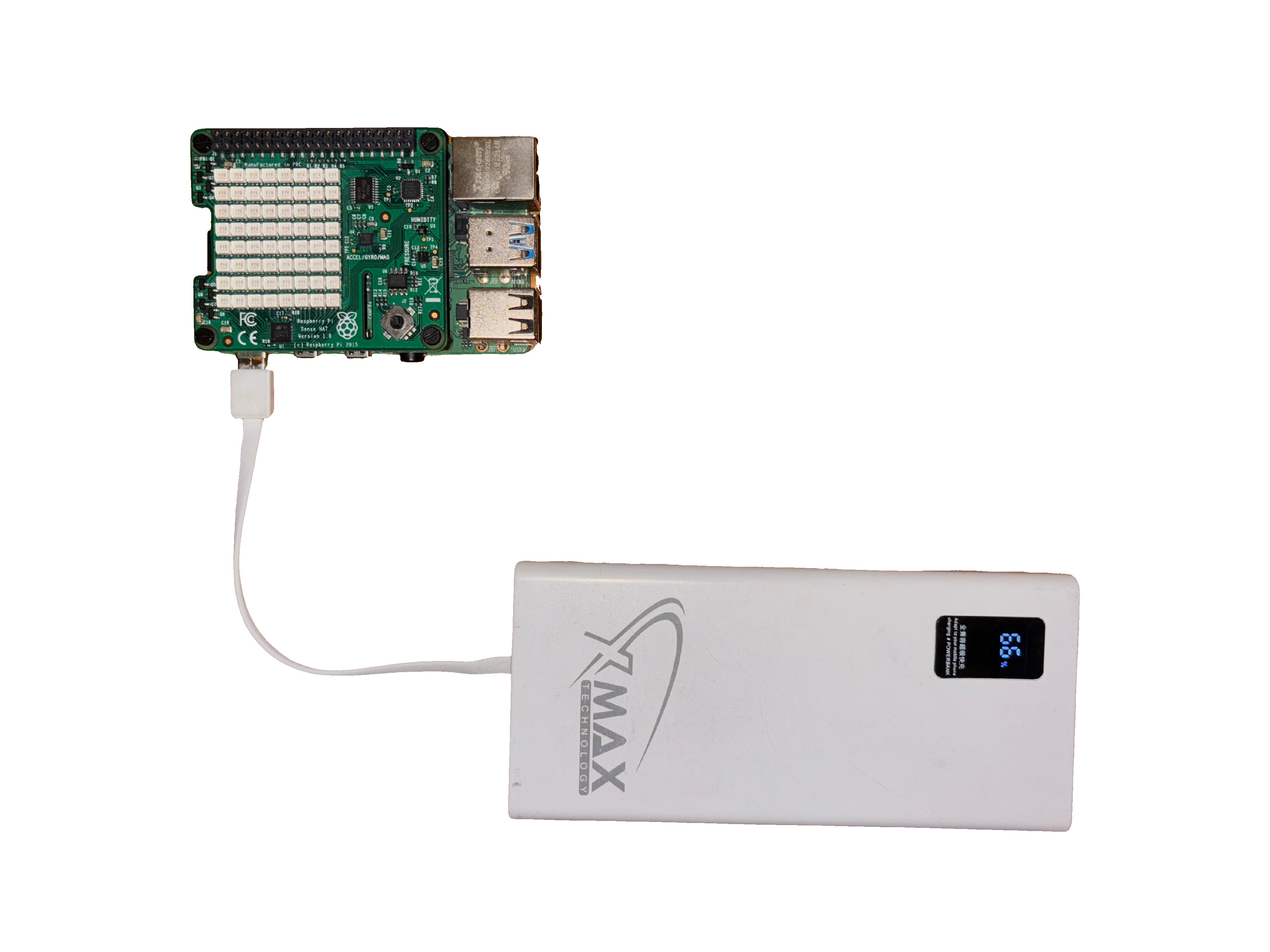}
    \caption{AmBox Node hardware setup}
    \label{fig:node-setup}
\end{figure}

The \emph{AmBox Node} is implemented using a Raspberry Pi 4 Model B equipped with an ARM 64-bit CPU, 8\,GB of RAM, and a 256\,GB microSD card, running Raspberry Pi OS. 
It acts as the primary control unit of the system, responsible for sensor data acquisition, local processing and storage, and interaction with the blockchain network. 
The physical setup of the Node is shown in Figure~\ref{fig:node-setup}.

Environmental data is collected using a Sense HAT connected via the GPIO interface. 
The Sense HAT integrates the following sensors:

\begin{itemize}
    \item Temperature sensor with a measurement range of \(0^\circ\mathrm{C}\) to \(65^\circ\mathrm{C}\) (±\(2^\circ\mathrm{C}\));
    \item Relative humidity sensor with a measurement range of \(0\%\) to \(100\%\);
    \item Barometric pressure sensor with a measurement range of \(260\,\mathrm{hPa}\) to \(1260\,\mathrm{hPa}\).
\end{itemize}

For portable deployment and testing, the Node is powered by a 10\,000\,mAh rechargeable lithium-polymer battery pack, enabling extended operation in environments without continuous power availability.

\subsubsection{AmBox Mote}

\begin{figure}[!htb]
    \centering
    \includegraphics[width=0.7\linewidth]{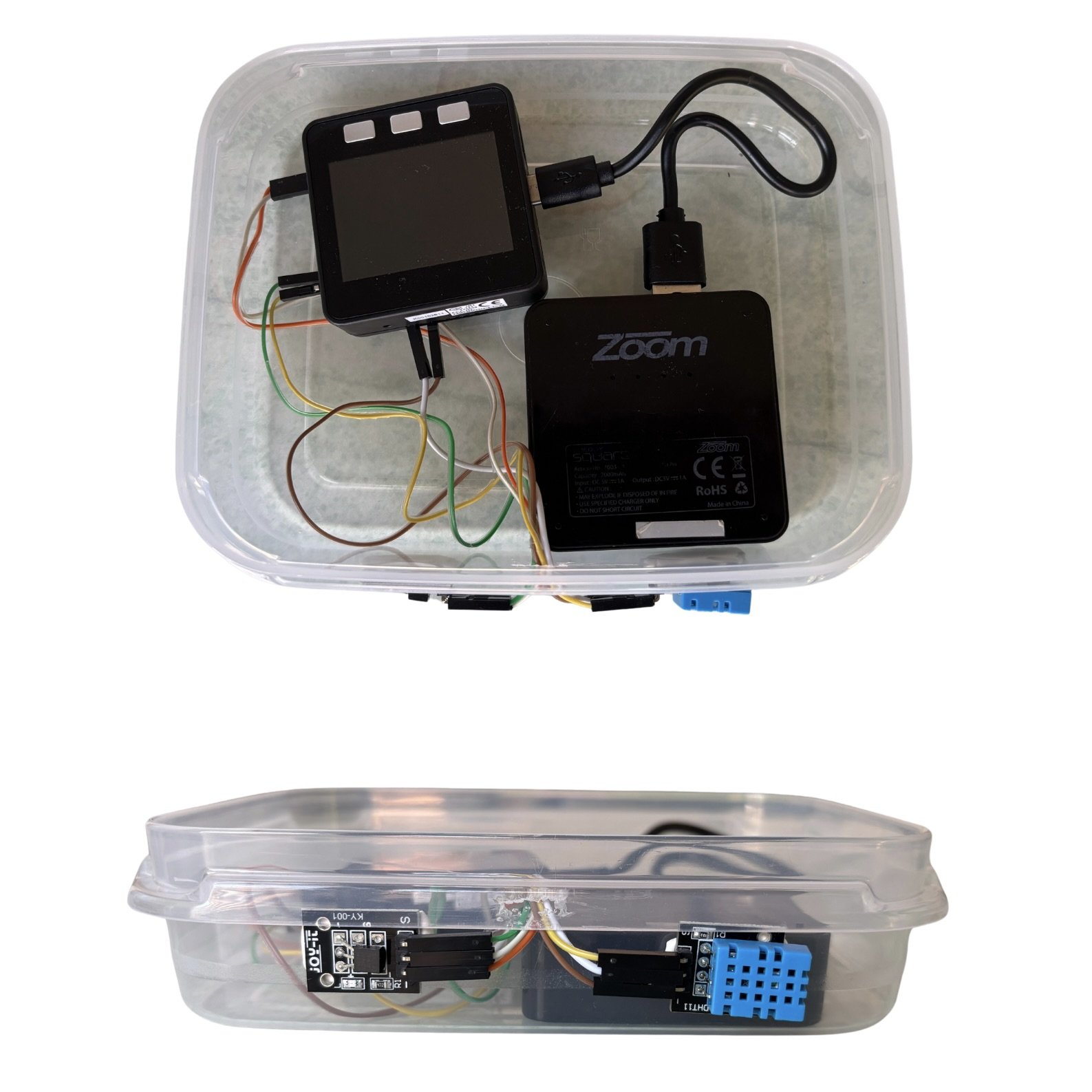}
    \caption{AmBox Mote hardware setup (top and side views)}
    \label{fig:mote-setup}
\end{figure}

The \emph{AmBox Mote} is implemented using an ESP32-based M5Stack Core microcontroller, featuring 520\,KB of SRAM and 16\,MB of flash memory. 
A 16\,GB microSD card is used for local buffering of collected sensor data. 
The Mote hardware setup is shown in Figure~\ref{fig:mote-setup}.

Additional sensors are connected via a breadboard using I\textsuperscript{2}C-compatible interfaces. 
The deployed sensors include:

\begin{itemize}
    \item KY-001 temperature sensor (DS18B20) with a measurement range of \(-55^\circ\mathrm{C}\) to \(125^\circ\mathrm{C}\) (±\(0.5^\circ\mathrm{C}\));
    \item KY-015 humidity sensor (DHT11) with a measurement range of \(20\%\) to \(90\%\) (±\(5\%\)).
\end{itemize}

Similar to the Node, the Mote is powered using a 10\,000\,mAh lithium-polymer battery bank, supporting mobile and energy-constrained deployments.

\subsubsection{Operator Server and Blockchain Network}

For prototyping and experimental evaluation, both the Operator Server and the Blockchain Network were hosted locally on a laptop equipped with an Apple M1 processor and 8\,GB of RAM. 
The Operator Server is implemented using the Express.js framework and exposes RESTful APIs for device management and monitoring.

The blockchain network is deployed using Docker containers and configured with Hyperledger Fabric v2.5, including the Fabric CA, orderer, peers, and CouchDB state database.

\subsection{Software Setup}

This subsection describes the software environments and configurations used to implement the \emph{AmBox} system components.

\subsubsection{AmBox Node}

The \emph{AmBox Node} software is primarily implemented in TypeScript, with auxiliary components written in Python for sensor interfacing. 
The main application runs on the Node.js runtime and exposes HTTP endpoints using the Express.js framework, enabling remote configuration, commissioning, and state control.

Upon startup, the Node can operate in one of three states:

\begin{itemize}
    \item \textbf{Idle state}: the Node is inactive and awaits instructions;
    \item \textbf{Heartbeat state}: the Node periodically reports its operational status without collecting sensor data;
    \item \textbf{Monitoring state}: the Node actively collects sensor data at configured intervals and attempts to report it to the blockchain.
\end{itemize}

State transitions are triggered via HTTP requests. 
The last operational state and configuration parameters are persisted in a local JSON file (\texttt{amboxConfig.json}), allowing the Node to restore its previous state after a reboot and resume interrupted monitoring tasks.

\subsubsection{AmBox Mote}

The \emph{AmBox Mote} is implemented using the Arduino framework in C++ and compiled using the Arduino toolchain. 
It operates exclusively as a sensing peripheral and does not communicate directly with the blockchain or the Operator Server.

The Mote communicates with a designated Node via Bluetooth Low Energy (BLE). 
Libraries such as \texttt{M5Stack.h}, \texttt{DallasTemperature.h}, and \texttt{BLEDevice.h} are used to interface with hardware components and manage BLE communication. 
Device behavior is implemented using BLE characteristic callbacks. 
Configuration parameters are persisted in a local \texttt{config.json} file, and collected data is buffered locally before transmission to the Node.

\subsubsection{Blockchain Implementation}

The \emph{AmBox} system uses Hyperledger Fabric to provide secure, decentralized, and tamper-evident storage of environmental monitoring data. 
The Fabric deployment includes a Certificate Authority, one orderer, and two organizations, each hosting a single peer node backed by CouchDB. Chaincode is implemented in TypeScript and deployed in a containerized environment using Node.js v18.20.7. \emph{AmBox Nodes} communicate with Fabric peers via gRPC using the \texttt{@hyperledger/fabric-gateway} and \texttt{@grpc/grpc-js} libraries. Each transaction submitted by an \emph{AmBox Node} includes a digital signature generated locally by the device. Signatures are verified on-chain using public key cryptography, ensuring data integrity, authenticity, and non-repudiation.

The chaincode defines an asset model inspired by the EPCIS standard\footnote{https://vb.nweurope.eu/media/14881/epcis-standard-12-r-2016-09-29.pdf}, where each asset represents an environmental report. 
Assets include metadata such as device identifier, product identifier, batch number, timestamp, and sensor readings, serialized in JSON format. 
The chaincode supports asset creation, querying, and retrieval of recent records.

\subsection{Communication}

The \emph{AmBox} prototype employs Wi-Fi and BLE as its primary communication technologies, balancing availability, range, and energy efficiency.

\subsubsection{Node--Mote Communication}

Communication between the \emph{AmBox Node} and \emph{AmBox Motes} is implemented using BLE. 
Motes operate as BLE peripherals that advertise sensor data, while the Node acts as a central device that scans for and connects to authorized Motes. Pairing is configured manually by sharing predefined BLE identifiers and the Node's MAC address with the Mote, reducing the likelihood of unintended connections in dense environments. Once connected, the Node subscribes to BLE characteristics and receives sensor updates via notifications, following a publish-subscribe model that avoids active polling. 

The Node implements automatic reconnection logic to detect disconnections and re-establish links when Motes return within range, ensuring robustness under mobility and intermittent connectivity.

\subsubsection{Node--Blockchain Communication}

Communication between the \emph{AmBox Node} and the blockchain is handled via the Hyperledger Fabric Gateway SDK for Node.js over secure gRPC channels protected by TLS. 
The Node loads its X.509 certificate and private key from the local file system and validates peer identities using the Fabric CA root certificate.

Sensor data buffered locally is periodically read, digitally signed, and submitted to the blockchain by invoking the \texttt{AddEvents} chaincode function. 
Submissions are serialized using a mutex to prevent concurrent file access. 
If a submission fails due to network or endorsement issues, the data remains stored locally and is retried in subsequent reporting cycles.

Each blockchain transaction follows these steps:

\begin{enumerate}
    \item Collection and local buffering of sensor data from the Node and associated Motes;
    \item Digital signing of the data and generation of a unique identifier;
    \item Submission of the signed data to the blockchain via a chaincode invocation;
    \item Endorsement and commitment of the transaction to the ledger;
    \item Retention and retry of data in case of submission failure.
\end{enumerate}

\subsubsection{Node--Operator Server Communication}

Communication between the \emph{AmBox Node} and the Operator Server is implemented via a RESTful API over HTTPS. 
This interface supports remote configuration, commissioning, monitoring control, and heartbeat-based status reporting. Table~\ref{tab:node-endpoints} summarizes the API endpoints exposed by the Node.

\begin{table*}[h]
    \centering
    \caption{AmBox Node API Endpoints}
    \label{tab:node-endpoints}
    \begin{tabular}{|l|p{5cm}|p{6cm}|}
        \hline
        \textbf{Endpoint} & \textbf{Parameters} & \textbf{Description} \\
        \hline
        /init & None & Starts the heartbeat service. \\
        /configHeartbeat & \texttt{ipaddr}, \texttt{port}, \texttt{heartbeat\_timeout} & Configures heartbeat parameters. \\
        /configBlockchain & \texttt{ipaddr}, \texttt{port}, \texttt{channel\_name}, \texttt{chaincode\_name} & Configures blockchain connection. \\
        /turnOff & None & Stops heartbeat and powers down. \\
        /startMonitoring & \texttt{prod\_id}, \texttt{batch\_no}, \texttt{interval}, \texttt{sensor\_params} & Starts monitoring. \\
        /stopMonitoring & None & Stops monitoring and resumes heartbeat. \\
        \hline
    \end{tabular}
\end{table*}

\begin{figure}[!htb]
    \centering
    \includegraphics[width=0.8\linewidth]{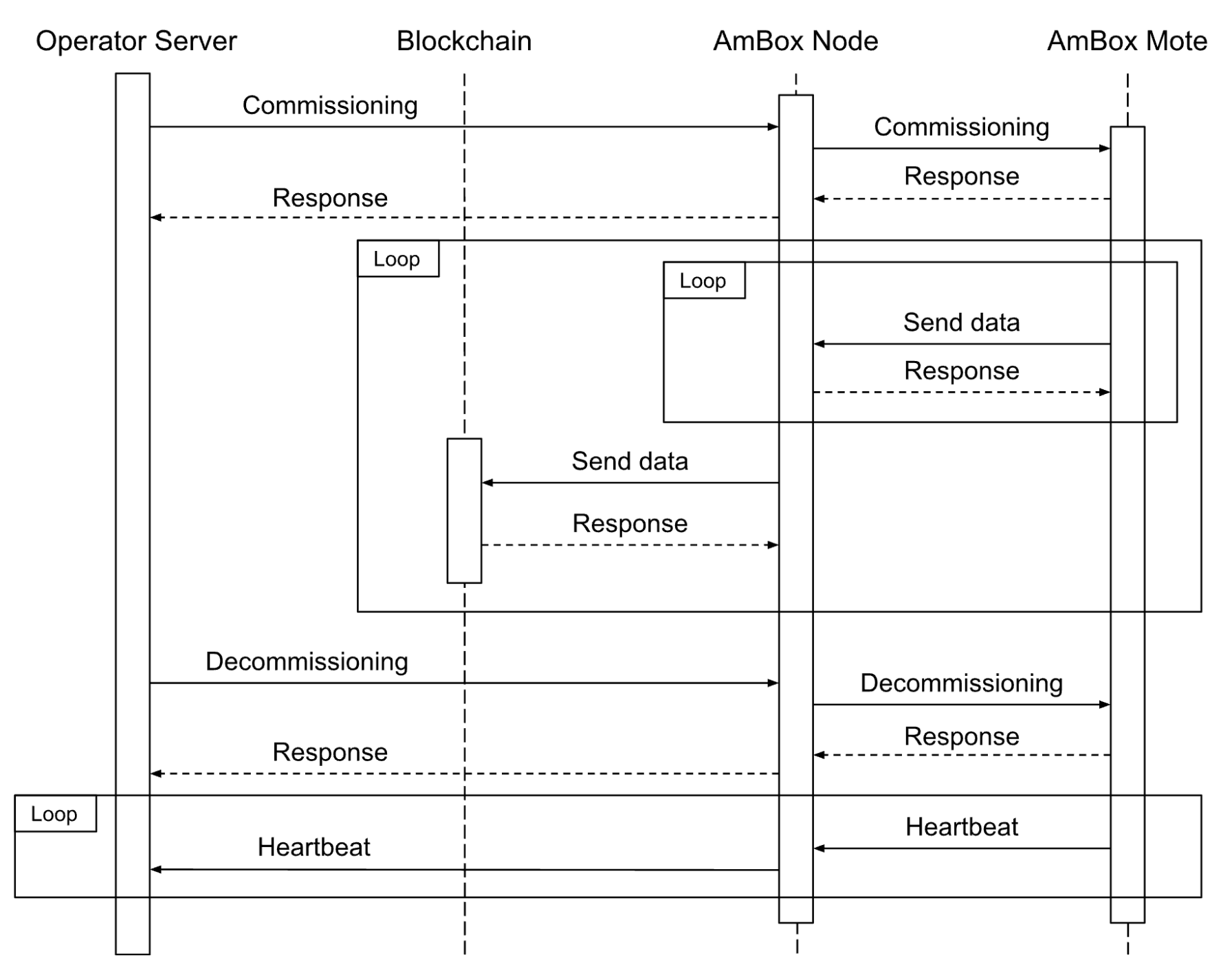}
    \caption{AmBox communication sequence}
    \label{fig:communications-sequence}
\end{figure}

Figure~\ref{fig:communications-sequence} illustrates the overall communication sequence. 
After commissioning, the Node propagates configuration parameters to the Motes, which begin data collection. 
Upon decommissioning, monitoring stops, remaining data is transmitted, and the system transitions to heartbeat mode.

\section{Evaluation}

The \emph{AmBox} system was evaluated in two complementary phases: (i) scenario-based validation and (ii) quantitative performance analysis. 
The first phase focused on validating core functional requirements, including connectivity resilience, data integrity, and secure synchronization, using both single-device and multi-device deployments. 
The second phase assessed system performance through controlled experiments measuring sensor accuracy, communication latency, and energy consumption.

\subsection{Phase 1: Scenario-Based Validation}

Two deployment configurations were evaluated: a standalone setup using a single \emph{AmBox Node} and a distributed setup combining a Node with one or more \emph{AmBox Motes}.

\subsubsection{Setup 1: Single-Device Deployment}

This setup evaluates the \emph{AmBox Node} operating independently. 
The Node collected environmental data (temperature, humidity, and pressure) and transmitted signed records directly to the blockchain. 
The experiment assessed correct blockchain integration, resilience to network disruptions, and enforcement of data integrity.

The Node was deployed with a local Wi-Fi connection and connected to a Hyperledger Fabric blockchain. 
It was configured to sample sensor data at regular intervals and submit aggregated records every 5 minutes. 
Under stable connectivity, data was successfully collected, signed, and committed to the blockchain without errors.

To evaluate robustness under unstable network conditions, the Node’s Wi-Fi connection was intentionally interrupted three times, for durations of 2 minutes, 15 minutes, and 1 hour. 
During each disconnection period, the Node continued to collect sensor data and buffered it locally in its JSON-based storage.

Upon restoration of connectivity, the Node automatically resumed communication with the blockchain and transmitted all buffered records in the correct temporal order. 
No data loss or corruption was observed, and timestamps confirmed that records were preserved exactly as collected. 
These results demonstrate that the local buffering and deferred submission mechanisms effectively provide resilience against temporary network outages.

To validate data integrity guarantees, tampering tests were conducted on locally stored records. 
After collecting and digitally signing multiple records, specific fields (e.g., temperature values) were manually altered in the local buffer prior to blockchain submission. 
In all cases, the chaincode detected the inconsistencies during signature verification and rejected the tampered records. 
This confirms that \emph{AmBox} reliably enforces data integrity through cryptographic signatures.

\subsubsection{Setup 2: Multi-Device Deployment}

This setup evaluates \emph{AmBox} in a distributed configuration, where an \emph{AmBox Mote} communicates with an \emph{AmBox Node} via BLE. 
The objective was to assess Node–Mote interaction, BLE reliability, and end-to-end data integrity.

The Mote sampled environmental data, digitally signed the measurements, and transmitted them to the Node over BLE. 
The Node aggregated the received data and forwarded it to the blockchain. 
This configuration enabled extended spatial coverage and validated multi-device synchronization.

To assess resilience to local communication failures, BLE connectivity between the Node and Mote was intentionally disrupted by physically separating the devices. 
Disconnections were introduced for 2 minutes, 15 minutes, and 1 hour, mirroring the durations used in the single-device setup. 
During each disruption, the Mote continued sampling data and buffered records locally.

Once proximity was restored, the Node automatically reconnected to the Mote and retrieved all buffered records. 
All recovered data was subsequently submitted to the blockchain, with no losses, duplication, or ordering issues observed. 
These results demonstrate that \emph{AmBox} tolerates temporary BLE failures and guarantees eventual data consistency through local caching and deferred synchronization.

As in Setup~1, integrity tests were conducted by altering signed records after collection. 
Any modification to data, whether during BLE transmission or while buffered on the Node, was detected during on-chain signature verification. 
This confirms that \emph{AmBox} provides secure, end-to-end data integrity across distributed sensing devices.

\subsection{Phase 2: Performance Analysis}

This phase evaluates the performance characteristics of \emph{AmBox} in terms of sensing accuracy, communication latency, and energy consumption.

\subsubsection{Sensor Accuracy}

\begin{figure}[!htb]
    \centering
    \includegraphics[width=0.9\columnwidth]{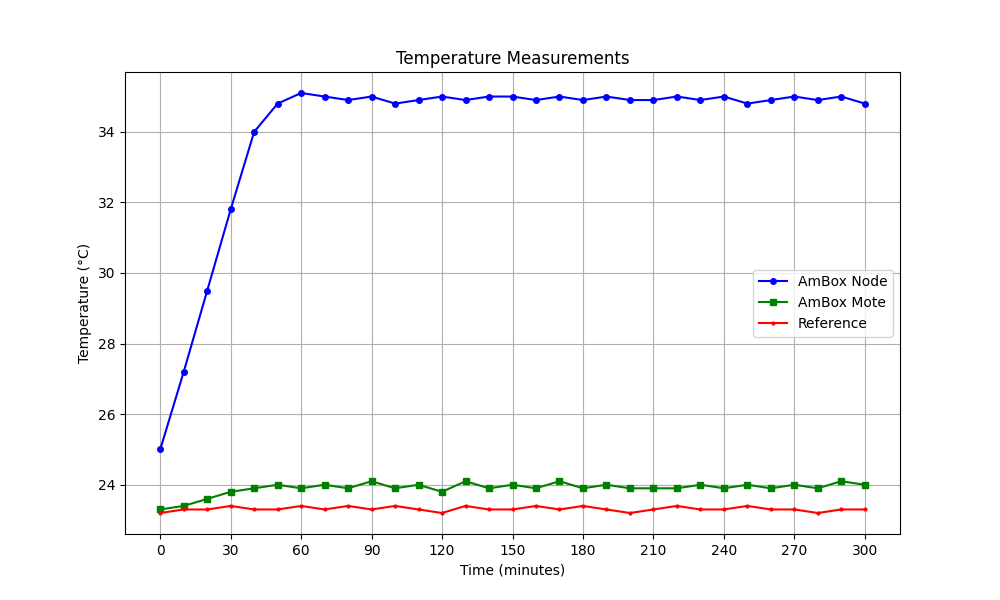}
    \caption{Temperature measurements over a 5-hour period}
    \label{fig:temperature-values}
\end{figure}

Sensor readings collected by the \emph{AmBox Node} and \emph{AmBox Mote} were compared against calibrated reference sensors over a 5-hour period. 
Figure~\ref{fig:temperature-values} shows the temperature measurements.

The Node consistently reported higher temperature values than the reference sensors, which can be attributed to thermal interference from the Raspberry Pi’s CPU. In contrast, the Mote’s readings closely matched the reference values, benefiting from externally connected sensors and lower internal heat generation. Similar trends were observed for humidity measurements, where the Node reported slightly lower values, likely influenced by temperature effects. Pressure readings were consistent across devices, with only minor calibration differences.

\subsubsection{Communication Performance}

Communication latency was evaluated for two paths: Node-to-blockchain communication over Wi-Fi and Mote-to-Node communication over BLE. Latency was measured using round-trip time (RTT), with 40 message exchanges per scenario. Table~\ref{tab:bc_latency} reports blockchain communication latency under local and remote deployment conditions. 
The local setup achieved an average latency of 148\,ms, while the remote setup exhibited higher latency, averaging 628\,ms, reflecting additional network overhead.

\begin{table}[htb]
\centering
\caption{Node-to-blockchain communication latency}
\label{tab:bc_latency}
\begin{tabular}{|c|c|c|c|}
\hline
\textbf{Network} & \textbf{Average (ms)} & \textbf{Min (ms)} & \textbf{Max (ms)} \\ \hline
Local  & 148.07 & 70  & 803  \\ \hline
Remote & 627.94 & 343 & 1992 \\ \hline
\end{tabular}
\end{table}

BLE communication between the Node and Mote exhibited significantly lower latency, with an average of 46\,ms, as shown in Table~\ref{tab:mote_latency}. 
This confirms the efficiency of local short-range communication for multi-device deployments.

\begin{table}[htb]
\centering
\caption{Node–Mote BLE communication latency}
\label{tab:mote_latency}
\begin{tabular}{|c|c|c|}
\hline
\textbf{Average (ms)} & \textbf{Min (ms)} & \textbf{Max (ms)} \\ \hline
46.07 & 27 & 109 \\ \hline
\end{tabular}
\end{table}

\subsubsection{Energy Consumption}

\begin{figure}[!htb]
    \centering
    \includegraphics[width=0.9\columnwidth]{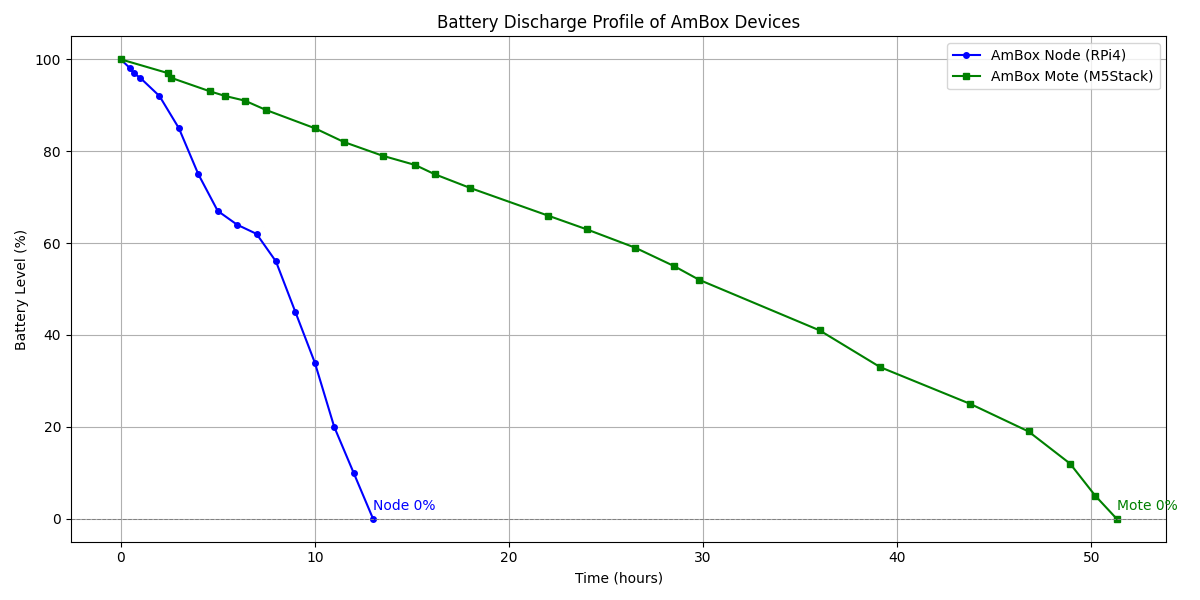}
    \caption{Battery discharge profiles of AmBox Node and Mote}
    \label{fig:battery-values}
\end{figure}

Energy consumption was evaluated using a single 10\,000\,mAh battery under typical operating conditions. 
Both devices sampled all available sensors once per minute and submitted data to the blockchain every 10 minutes. 
Figure~\ref{fig:battery-values} illustrates the battery discharge profiles.

The \emph{AmBox Node} operated for approximately 13 hours, while the \emph{AmBox Mote} operated for approximately 51 hours before battery depletion. 
Both devices satisfy the minimum requirement of 10 hours of autonomous operation. 
The significantly longer battery life of the Mote reflects its more energy-efficient hardware design, while the Node’s consumption highlights potential optimization opportunities for future iterations.

\subsection{Transport Tracking Case Study}

In addition to controlled experiments, a real-world case study was conducted during a 4-hour bus trip from Loulé to Lisbon. 
An \emph{AmBox Node}, powered by a 10\,000\,mAh battery and placed inside a suitcase, collected environmental data every 5 minutes and attempted to upload records to the blockchain. Due to limited Wi-Fi connectivity during the journey, the system remained offline for most of the trip. 
Nevertheless, all data was collected and buffered locally, and successfully uploaded to the blockchain once connectivity was restored. 
After approximately 5 hours of operation, the battery level remained at 69\%, indicating moderate power consumption suitable for extended transport scenarios.

Temperature readings were higher than expected, primarily due to thermal interference from the Raspberry Pi CPU and the enclosed deployment environment. 
This limitation highlights the need for improved thermal isolation or external sensor placement in future hardware revisions.

Overall, the case study confirms that \emph{AmBox} can reliably operate under real-world mobility and connectivity constraints, preserving data integrity and ensuring eventual synchronization with the blockchain.

\section{Discussion}

The evaluation of the \emph{AmBox} prototype highlights both the strengths of the proposed system and several areas for improvement that inform future design iterations. One of the most notable limitations observed concerns temperature measurement accuracy on the \emph{AmBox Node}. The proximity of the temperature sensor to the Raspberry Pi’s CPU resulted in consistently higher readings compared to calibrated reference sensors, particularly when the device was deployed in an enclosed environment such as a suitcase. 
This indicates that thermal interference from the processing unit can significantly affect sensor accuracy, and suggests that similar effects may influence other onboard sensors. 
Future versions of the Node should therefore consider improved thermal isolation, external sensor placement, or hardware-level compensation techniques to mitigate heat-related measurement bias.

Communication experiments demonstrated that the system performs reliably under local network conditions, while higher latency was observed when interacting with a remote blockchain network. 
This behavior is expected due to increased network distance and variability, and reinforces the importance of the local buffering and deferred synchronization mechanisms implemented in \emph{AmBox}. 
These mechanisms enable uninterrupted data collection during connectivity outages and ensure eventual consistency once network access is restored, which is essential for mobile and geographically distributed supply chain deployments.

In terms of energy consumption, both the Node and the Mote met the minimum operational requirement of 10 hours of autonomous operation, confirming their suitability for field use. 
The \emph{AmBox Mote} exhibited significantly longer battery life than the Node, reflecting its lower-power hardware design and simpler processing tasks. 
While the Node’s shorter battery life remains acceptable, it highlights an opportunity for optimization through hardware selection, duty-cycling strategies, or offloading non-critical tasks to reduce energy consumption in future revisions.

The system also demonstrated strong resilience in maintaining data integrity under intermittent connectivity conditions. 
Sensor data was consistently buffered locally during offline periods and successfully synchronized with the blockchain once connectivity was restored, without loss or duplication. 
Similarly, the Mote tolerated periods of Bluetooth unavailability and reliably recovered buffered data, confirming the effectiveness of the system’s fault-tolerant design.

Finally, the evaluation confirmed that \emph{AmBox} can detect unauthorized data modifications through cryptographic signature verification, providing a robust baseline for data integrity and trust. 
However, the security evaluation was limited to basic tampering scenarios. 
More sophisticated attack vectors, such as key compromise, replay attacks, or denial-of-service scenarios, were not considered in this prototype. 
Addressing these threats through extended threat modeling, stronger key management, and additional security controls represents an important direction for future work.

\section{Conclusion}

This paper presented \emph{AmBox}, a modular and low-cost system for blockchain-backed environmental monitoring in agri-food supply chains. 
By integrating IoT sensing devices directly with a permissioned blockchain, \emph{AmBox} enables tamper-evident and verifiable recording of ambient conditions such as temperature and humidity, without reliance on centralized intermediaries.

Two deployment configurations, a standalone Node and a distributed Node--Mote architecture, were designed and evaluated, demonstrating flexibility across fixed and mobile supply chain scenarios. 
Experimental results confirmed the system’s ability to operate under intermittent connectivity, preserve data integrity, and reliably synchronize monitoring data with a Hyperledger Fabric blockchain.

Future work will focus on improving sensor accuracy and energy efficiency, strengthening security through hardware-based protections, and enhancing scalability and integration with existing supply chain management systems. 
These developments will further support reliable, end-to-end traceability in real-world agri-food deployments.

\section*{Acknowledgements}

Work supported by national funds through Fundação para a Ciência e a Tecnologia, I.P. (FCT) under projects UID/50021/2025 (DOI: \url{https://doi.org/10.54499/UID/50021/2025}) and UID/PRR/50021/2025 (DOI: \url{https://doi.org/10.54499/UID/PRR/50021/2025}) and by Blockchain.PT – Decentralize Portugal with Blockchain Agenda, (Project no 51), WP 1: Agriculture and Agri-food, Call no 02/C05-i01.01/2022, funded by the Portuguese Recovery and Resilience Program (PRR), The Portuguese Republic and The European Union (EU) under the framework of Next Generation EU Program.

\bibliographystyle{unsrtnat}
\bibliography{paper}

\end{document}